\documentclass[conference]{IEEEtran}

\usepackage{amsmath}
\usepackage{amssymb}
\usepackage{color}
\usepackage{array}
\usepackage{multirow}
\usepackage{multicol}
\usepackage{caption}
\usepackage{subcaption}
\usepackage{float}
\usepackage{algpseudocode}
\usepackage{breqn}

\usepackage[ruled]{algorithm2e} 
\usepackage{changepage}
\usepackage[table]{xcolor}

\usepackage[utf8x]{inputenc}

\usepackage{textcomp,marvosym}

\usepackage{cite}

\usepackage{nameref,hyperref}
\usepackage{graphicx}

\ifCLASSINFOpdf
\else
\fi

\begin{document}
%
\title{Can social influence be exploited to compromise security: An online experimental evaluation}

\author{\IEEEauthorblockN{Soumajyoti Sarkar, Paulo Shakarian}
\IEEEauthorblockA{Arizona State University, Tempe, USA\\
Email: \{ssarka18, aleali1, shak\}@asu.edu}
\and
\IEEEauthorblockN{Mika Armenta, Danielle Sanchez, Kiran Lakkaraju}
\IEEEauthorblockA{Sandia National Laboratory, New Mexico, USA\\
Email: \{mlarmen, dnsanc, klakkar\}@sandia.gov}}


%


\maketitle

\begin{abstract}

Social media has enabled users and organizations to obtain information about technology usage like software usage and even security feature usage. However, on the dark side it has also allowed an adversary to potentially exploit the users in a manner to either obtain information from them or influence them towards decisions that might have malicious settings or intents.  While there have been substantial efforts into understanding how social influence affects one’s likelihood to adopt a security technology, especially its correlation with the number of friends adopting the same technology, in this study we investigate whether peer influence can dictate what users decide over and above their own knowledge.  To this end, we manipulate social signal exposure in an online controlled experiment with human participants to investigate whether social influence can be harnessed in a negative way to steer users towards harmful security choices.  We analyze this through a controlled game where each participant selects one option when presented with six security technologies with differing utilities, with one choice having the most utility. Over multiple rounds of the game, we observe that social influence as a tool can be quite powerful in manipulating a user's decision towards adoption of security technologies that are less efficient. However, what stands out more in the process is that the manner in which a user receives social signals from its peers decides the extent to which social influence can be successful in changing a user's behavior.
\end{abstract}

\section{Introduction}

Social influence is key to technology adoption, and there have been a lot of recent attention on understanding the role of persuasion in security technology adoption \cite{sec_sauvik, soc_inf_sec}. These studies have focused on various social influence factors that impact a user's decision when adopting several security features.
However, the question that has  evaded  the purview of social influence for security adoption is its role in the era of cyber-adversaries ans especially when their intents cam be as malicious as attempts to hack into systems. Specifically, social influence has always been studied from the perspective of having a net positive impact on society \cite{sauvik_2}, especially when considering their utility in behavior diffusion.

However, the occurrence of strategic events in the past in which cyber warriors exploit social media  with malicious intents has opened new avenues for reconsidering influence as a tool for change. Consider the example of the experiment where an American security firm created fake Facebook accounts of a fictitious user in order to entice users to befriend her and inappropriately share information \cite{shakarian}. In the course of the experiment, the study showed that the role of transitive trust factored into influencing users to make connections with her and in some cases even share sensitive geo-location information, especially as users did not verify the account.
Specifically, some of the questions that emerged out of such studies involved understanding whether users place too much trust on friends and too little on their acquired knowledge when adopting security technologies. And then from the perspective of a cyber adversary, can social influence be ``successfully'' leveraged for evil persuasion where users could be tricked into adopting technologies that might be less secure or could be used for hacking into their systems?

To this end, we conduct an online controlled experiment with human participants with the goal of understanding the impact of social influence on users when they decide to adopt security technologies. Specifically, we modulate the number of signals that users receive from their peers in the social network and the \textit{pattern} through which they receive them over time.  We analyze its effect through a game having multiple rounds. In every round, each participant makes a decision to select one security technology when presented with six choices with differing utilities, with one technology having the most utility. An example of such a pattern of peer influence in shown in Figure~\ref{fig:linear_cascade}. We investigate three main questions: (1) can users be influenced to deviate from the optimal security technology in presence of social signals, (2) does influence encourage users to explore more options despite having the knowledge of the optimal technology  and (3) does the role of social influence factor more than other aspects like exploration that are artifacts of the experimental setup. 

We observe from  our experiments that while an early exposure to higher social signals successfully influence users to deviate from the optimal technology, it is not able to retain the effect following the exploration period of the users. On the contrary, a delayed exposure to higher quantity of social signals influences individuals to deviate from the optimal choice at the end of the game. Additionally, we find that socially signals factor more than other cognitive aspects arising inherently from the design, such as the number of options explored by user so far or the number of alternating switches made.

\section{Methods}\label{sec:methods}
We ran an online, controlled decision-making game hosted by the Controlled Large Online Social Experimentation (CLOSE) platform and developed at Sandia National Laboratories \cite{lakkaraju2015controlled}, in which participants took on the role of a security officer at a bank. Participants were told that they and several of their peers at different banks were being asked to invest in a cyber-defense provider once a month for 18 months or rounds\footnote{We use Rounds/Months/Timesteps interchangeably but which refer to one discrete unit of time in our study}. All participants could view brief descriptions of provider capabilities – e.g. one of them being ``Secure.com utilizes algorithmic computer threat detection to keep systems safe. It prides itself on its efficiency and success rate in warding against attacks." Participants were able to choose from 6 different providers - among which only one was optimal, preventing 7 attacks. The remaining 5 providers prevented 6 attacks each (from that perspective, all suboptimal technologies had the same utility). This information about the optimal and suboptimal providers is not available to the participants in any group. An example of the screen is shown in Appendix\footnote{Due to space limitation Appendix is uploaded online: link}. We separated participants into 5 groups based on \textit{pattern} of social signal exposure which will be described in details in the $Design$ subsection following this. For each group, we controlled the number of signals (corresponding to a suboptimal technology) that were sent to an individual from their peers over each time step.  

The entire game was partitioned into two phases. For the first 12 rounds, no other information but a short excerpt about the six potential providers was given. After the participants made their selection for a given month, they saw the number of attacks their provider had prevented in the corresponding period. For every attack they prevented, participants received \$0.02. Thus, they were incentivized to avoid more attacks and earn more money. However, since the participants have to explore the technologies to first acquire the knowledge of the technology utilities, the first 12 rounds allow for individual decision making and exploration in the absence of any external knowledge about the technologies or peers. 

\begin{figure}[t!]
	\centering
	\includegraphics[width=8cm, height=2.3cm]{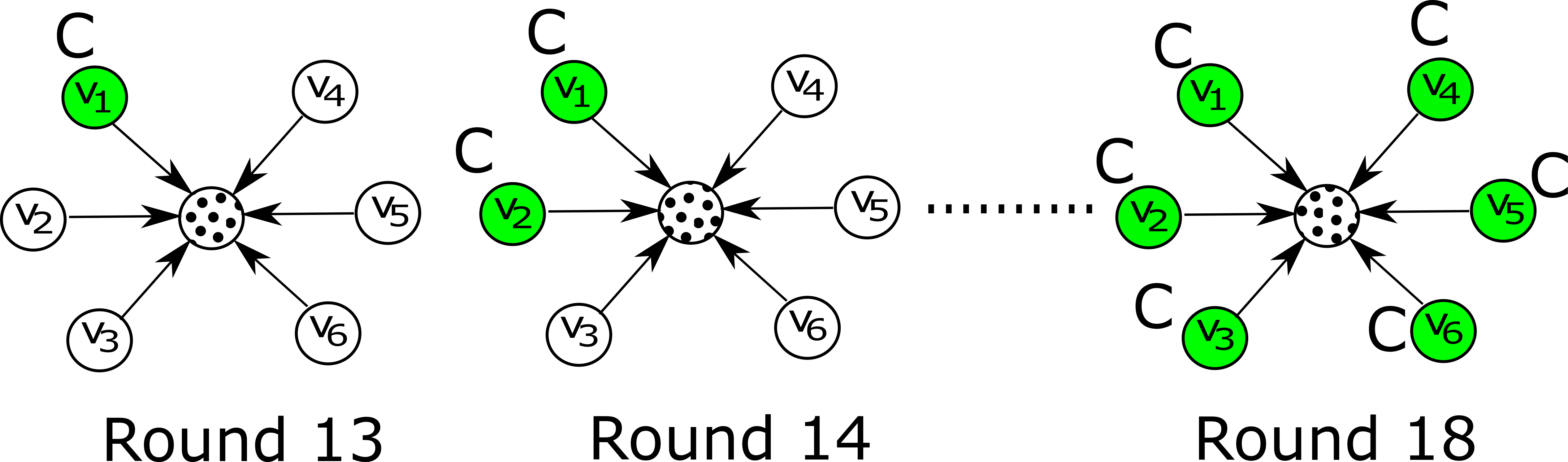}
	\caption{Illustration of the \emph{linear cascade} diffusion. The technology $C$ chosen by us as the suboptimal technology for influencing the user (in dots) cascades through the peers of the user over the 6 time steps. Colored nodes denote the activated peers w.r.t. $C$ (manually preprogrammed by us) at each time step. Note that although at Rounds starting at 13 and ending at 18, there are subjects (uncolored) among peers who have not adopted $C$, their decisions (technologies adopted which may not be $C$) at those time steps are visible to the subject in consideration (in dots). However, which users among the peers have been preprogrammed manually is by default unknown to the target subject. }
	\label{fig:linear_cascade}
\end{figure}

In the second phase of the experiment which started at Round 13, we introduced social influence by allowing participants to see their peers’ decisions after every round, and by varying the pattern by which concentric decisions among the peers were made over time. We attempt to avoid network effects by using pre-programmed bots (these are the peers that the users see in their screen) and holding the network structure constant. All individuals in all groups have 6 neighbors whose choices could be viewed by the corresponding individuals. An example of the network structure is shown in Figure~\ref{fig:linear_cascade}, where a participant receives social signals from its six neighbors about a random technology $C$ among the 5 providers (barring the provider with the optimal technology). For each subject, this suboptimal technology $C$ was selected as the peer choice that would be disproportionately signaled by peers over time (this pattern of influence would be manipulated by us). We call this $C$ the \textit{influence decision} for the respective user. The motivation behind this deliberate selection of suboptimal $C$ as the peer choice (controlled by us) is to investigate whether participants would be tempted to select the suboptimal choice.The knowledge of the peers' decisions allows a participant to rethink its own choices made in the presence of utilities and we use this to quantitatively measure the extent to which social influence is at play. We describe the signal patterns and conditions unique to each study in the following $Design$ section.

For all the research questions we investigate, the outcomes of interest are the decisions  made by participants in the last six rounds, in the presence of social signals from peers. 

\subsection{Participants}
We recruited a total of 357 participants for this study to play the same cyber-defense provider game.  Participants were paid \$2 with the opportunity to earn up to \$4.52 since as mentioned before,  they received a bonus of \$0.02 for every attack they prevented. Thus, the participants have a motivation to prevent more attacks in order to earn more money. 

\subsection{Design}
Participants were randomly assigned to five groups with each group having unique members not involved in decision making as part of other groups. Decisions in the first 12 rounds (first phase) are made without any peer signals and are equal between groups. Let $V(t)$ ($t$ denotes a time step or round among the last 6 rounds) denote the number of peers of a user who at time $t$ were programmed to select a chosen suboptimal technology by us. For the last six rounds participants in each group, except one group, receive signals in the following mechanism which we denote as the \textit{patterns of influence}: (1) \textit{No Message (NM)}: V = \{0, 0, 0, 0, 0, 0\}, (2) \textit{Uniform Message (UM) }: V = \{1, 1, 1, 1, 1, 1\}, (3) \textit{Linear Cascade (LC)}: V = \{1, 2, 3, 4, 5, 6\}, (4) \textit{Early Cascade (EC)}: V = \{4, 5, 6, 6, 6\} and, (5) \textit{Delayed Cascade (DC)}: V = \{1, 1, 1, 4, 5, 6\}.  (Figure~\ref{fig:signal_rate} shows the signal patterns for the groups):

Note that in all conditions, users can switch back to any choice in the next round after having selected an option in the current round. We consider the NM and UM groups as our baseline groups and LC, EC, DC groups as our treatments groups of interest. Note that while the pattern remains the same, the suboptimal technology chosen by us for each subject and that cascades through its peers is random. An example of the Linear Cascade (LC) pattern is shown in Figure~\ref{fig:linear_cascade}, where the subject (marked in dots) is able to receive social signals from its 6 neighbors. At Round 13 (start of the second phase), a signaler (node v1) selects a suboptimal provider C, and over the next five rounds, the remaining peers adopt the same behavior one after another.(we will refer to this as the influence decision). Note that although we program only a few of the peers (bots)  to adopt $C$ over time, the subjects are able to view all the bots and their decisions in their dashboard for all the last 6 timesteps.  We emphasize that all the peers of the subjects are bots and do not share any topology between themselves, thereby we sideline the effects of network on the individual behavior changes.

\section{Analysis} \label{sec:analyses}

\begin{table}[h]
	\begin{center}
		\begin{tabular}{|c|c|c|}
			\hline
			Group & $\#$ participants & Average number of attacks prevented \\\hline
			$NM$ & 55 & 105.2 \\\hline
			$UM$ & 71 & 106.28 \\\hline
			$LC$ & 79 & 103.81 \\\hline
			$DC$ & 81 & 104.8 \\\hline
			$EC$ & 71 &  103.83 \\\hline
		\end{tabular}
	\end{center}
	\caption{Average number of attacks prevented by subjects in each group. The lower attacks suggest participants deviated more from the optimal decision responding to social infleunce.}
	\label{tab:attacks_prev}
\end{table}
\vspace*{-.2in}
\subsection{ Distribution of attacks prevented}

Table~\ref{tab:attacks_prev} shows the distribution of attacks prevented by subjects in each group. We observe that, on average subjects in the EC and LC groups prevent more attacks compared to others.  Based on a survey analysis, we found that none of the traits like computer anxiety, computer confidence, computer liking, intuition or neuroticism  were correlated to the number of attacks prevented in all groups. \\

\noindent \textit{RQ1. Will participants deviate from the optimal security technology and move towards their peer suboptimal choice in the presence of social signals?} \\

Under this research question, we investigate two components. First, we try to investigate  whether the social signals prompt the users to deviate from the best security technology and move towards the suboptimal choice made by their peers - to this end, we first measure the proportional of individuals in the last step (Round 18) who do not opt for the optimal decision and the proportion who settle on their respective \textit{influence decision}. The second  component following this is to now measure these two metrics for each subject in each group at the time step (round) when the \textit{influence decision} programmed by us for a user is reflected in majority of the peers of that user. That is to say, the first time step when 4 out of 6 peers adopt the influence decision (> 50\% peers) - this happens at round 13 for EC and at round 16 for LC and DC groups.

\begin{table}[!h]
\begin{center}
\begin{tabular}{|l|l|l|l|l|l|}
\hline
\textbf{At the last step (Round 18)}                                  & \textbf{NM} & \textbf{UM} & \textbf{LC} & \textbf{DC} & \textbf{EC} \\ \hline
\begin{tabular}[c]{@{}l@{}}Proportional of individuals \\ not on optimal \end{tabular}       & 45.61           & 30.55           & 48.10           & 49.41           & 43.83           \\ \hline
\begin{tabular}[c]{@{}l@{}}Proportional of individuals \\ on influence decision \end{tabular} & 8.77           & 18.05           & 20.25           & 22.5           & 19.17          \\ \hline
\end{tabular}
\end{center}
\caption{}
\label{tab:tab_1}
\end{table}

For the first component, Table~\ref{tab:tab_1} shows the proportion of users in each group for the two metrics discussed above - the results show that the DC group participants deviated most from the optimal with 49.41\% of users settling on suboptimal choices in the last round. Among these, around 22.5\% of DC participants switched to their influence decision (which is different for each participant) which is also the maximum among all the groups. In fact while these results shed light on the retention power of the influence patterns - while the DC group's power of retention could be attributed to late exposure to larger quantity of peer signals, the EC group fails to retain a lot of the users after the initial rounds. 

\begin{table}[!h]
\begin{center}
\begin{tabular}{|l|l|l|l|}
\hline
\begin{tabular}[c]{@{}l@{}}\textbf{First step where majority} \\ \textbf{peers reflect influence} \\ \textbf{decision} \end{tabular}                                  & \begin{tabular}[c]{@{}l@{}} \textbf{LC} \\ \textbf{(Round 16)} \end{tabular} & \begin{tabular}[c]{@{}l@{}} \textbf{DC} \\ \textbf{(Round 16)} \end{tabular} & \begin{tabular}[c]{@{}l@{}} \textbf{EC} \\ \textbf{(Round 13)} \end{tabular} \\ \hline
\begin{tabular}[c]{@{}l@{}}Proportional of individuals \\ not on optimal \end{tabular}       &  46.83           & 54.11           & 57.49          \\ \hline
\begin{tabular}[c]{@{}l@{}}Proportional of individuals \\ on influence decision \end{tabular} & 16.45           & 24.70           & 30.13         \\ \hline
\end{tabular}
\end{center}
\caption{}
\label{tab:tab_2}
\end{table}

For the second component, Table~\ref{tab:tab_2} shows that on the contrary, the EC pattern of influence is able to draw more participants at the time step where the participant clearly observes its influence decision as the one that majority of its peers ($geq$4 out of 6) choose. This suggests that while early subjugation to exposures demonstrates a better proxy for social influence, the exploration time following this early exposures motivates users to move away from this decision and so we see a substantial drop in the values for the last step for EC (compared to DC and LC) from table~\ref{tab:tab_1}.  \\

\noindent \textit{RQ2. Does the presence of social signals influence users to explore and revisit different options?} \\

To further measure variations in decisions, we try to analyze the effect of the pattern of influence on decision explorations by users. The goal is to understand whether the introduction of peer signals prompts users to explore more options even in the presence of already acquired knowledge. 
Given a list of decisions $X$ made by each subject over the last 6 rounds, we define entropy as $H(X) = -\sum_{i=1}^{n} P(X_i) \log_b P(X_i)$ where $b$ is the base of the logarithm and $n$ denotes the number of possible decisions, which is 6 in our case. We use $b = 6$ to normalize the entropy values to be between 0 and 1 -- we note that this is an artifact of the experiment design as there are 6 decisions types. Intuitively, given decisions made by two subjects, the participant with higher entropy value has changed its decisions more frequently compared to a participant with lower entropy value.

\begin{figure}[t!]
	\centering
	\includegraphics[width=6cm, height=4cm]{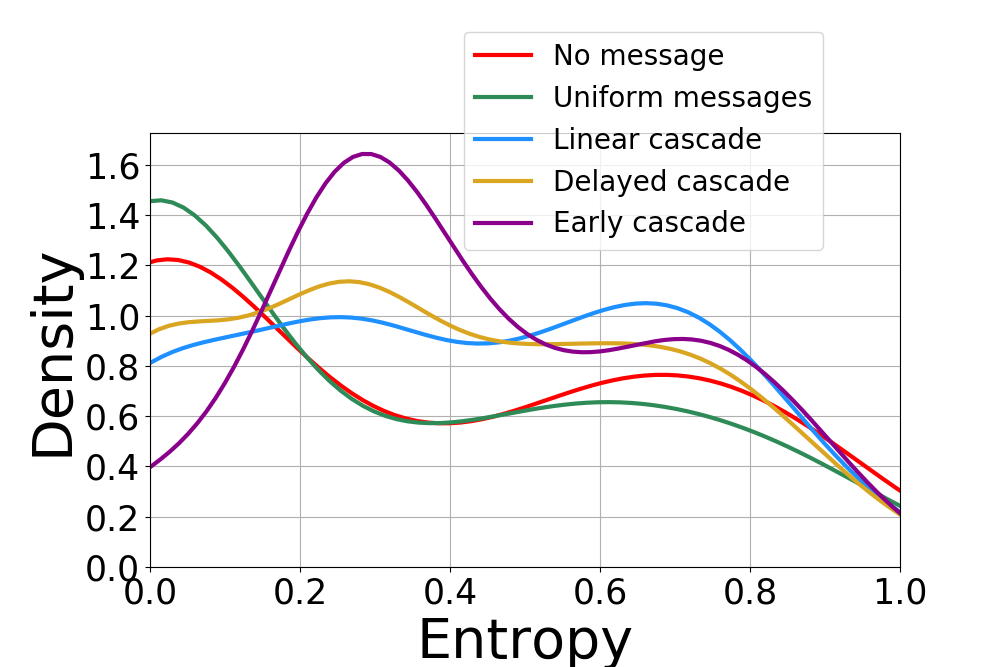}
	\caption{Entropy distributions for the second phase (Rounds 13 to 18).}
	\label{fig:ent}
\end{figure}

\figurename~\ref{fig:ent} shows the entropy of the decisions for the second phase of the experiment. We performed the Kolmogorov-Smirnov (KS) tests between pairwise distributions and we found that with respect to the UM and NM participants (control), the distributions showed statistically significant differences with the LC, EC and DC participants ($p$ $<$ 0.05 for all these pairwise tests) with respect to the options explored. Particularly, we find that for the control groups NM and UM, the entropy distributions peak near 0, which explains the fact that users do not explore much under the control setup whereas the peak for EC group is evident at around entropy of 0.3 suggesting that more users explore options in this group.

We conclude from this basic analysis that the introduction of peer influence in the form of treatment patterns of influence does indeed prompt users to explore more, but the more subtle question here is whether the exploration differs among the LC, EC and DC groups. To this end, we find that from the same pairwise KS tests, there is no statistically significant difference between these 3 groups when considered in pairs. However, while most of these distributions are multi-modal (having multiple peaks), the LC group tends to have more users having higher entropy shown by the observation that one of its modes lie near entropy value of 0.6. \\


\noindent \textit{RQ3. Does peer signals factor more than other cognitive aspects that might impact users like the number of switches made by the user so far?} \\

We use Cox proportional hazards model, which is the standard technique for assessing contagion in economics, marketing, and sociology \cite{survival}. This tool measures the hazard or likellihood of adoption of an individual at time $t$ as a function of individual characteristics and social influence: $\lambda(t, X_{ti})$ = $\lambda_{0t}exp(X_{ti}\beta)$ where $\lambda$ represents the hazard of adoption for a subject after the $t^{th}$ round ($t \in  Rounds[13, 18]$), $\lambda_{0t}$ represents the baseline hazard of adoption and $X_{ti}$ represents the static set of covariates  for subject $i$ after round $t$ - namely the number of signals reflecting \textit{influence decision}, the number of decision switches made by the participant at $t$ and the number of technology options (among the 6 possible) explored by the participant at $t$. 
\begin{figure}[!t]
	\centering
	\minipage{0.2\textwidth}
	\includegraphics[width=4.2cm, height=3cm]{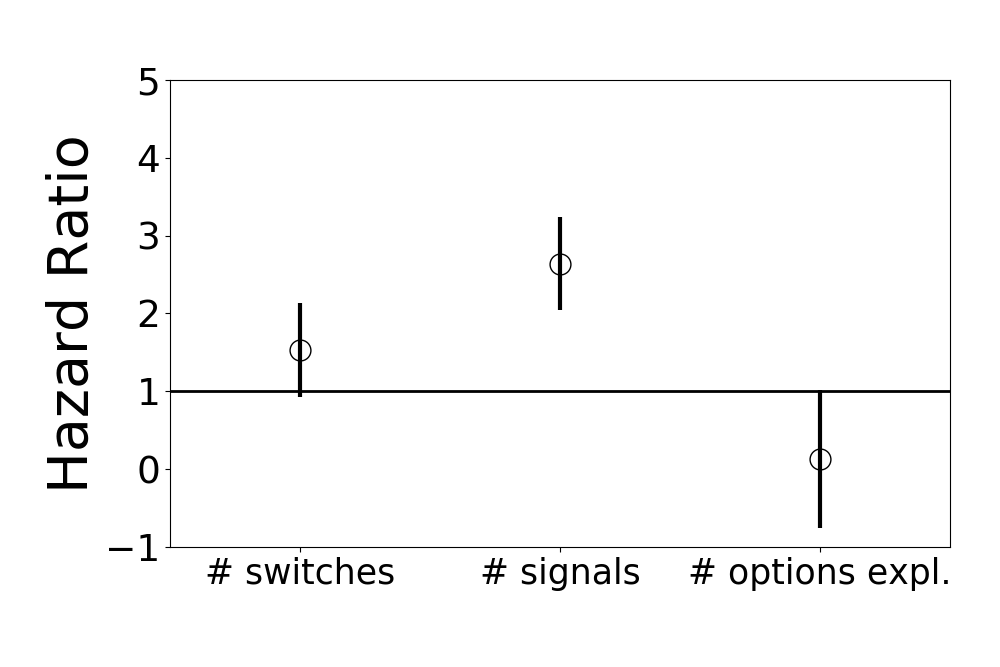}
	\subcaption{}
	\endminipage
	\hfill
	\minipage{0.23\textwidth}
	\includegraphics[width=4.2cm, height=3cm]{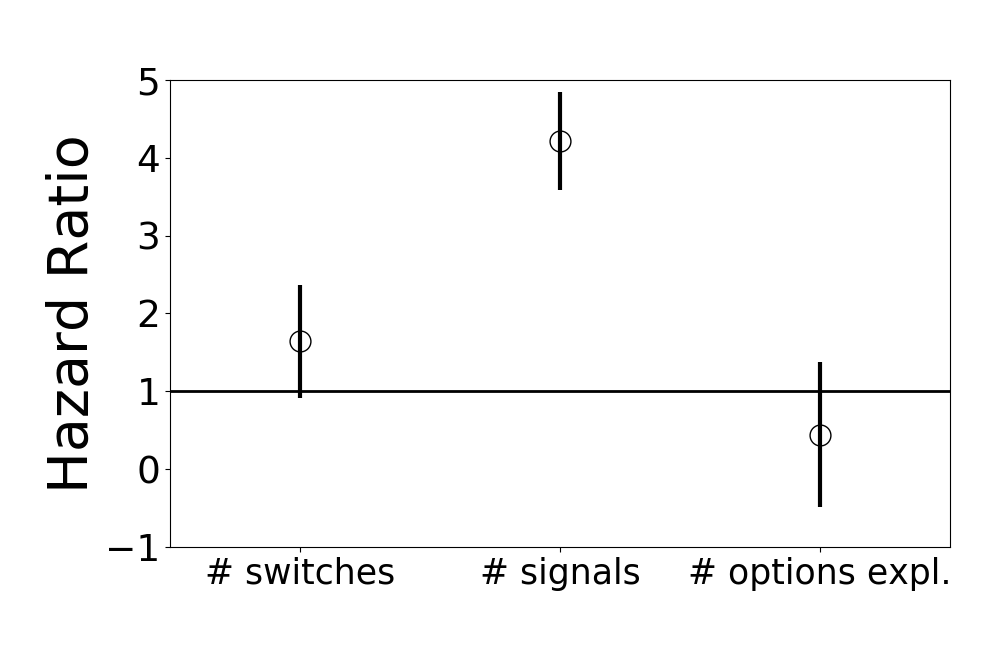}
	\subcaption{}
	\endminipage
	\hfill
	\caption{The signal vs time step plots for the 4 patterns - note that for the NM group (not shown here), no peer signal in the form of pre-selected suboptimal technologies were sent to the participants at any time step.}
	\label{fig:survival}
\end{figure}

Figure~\ref{fig:survival} displays the results for the DC and EC group participants and it shows that when the 3 factors are considered together for the hazard of adoption, the number of signals is the only significant factor playing into the adoption of the security features - the 95\% CI lies above 1 (which denotes significance) for both the DC and EC participants. However, we find that none of the factors were significant for the LC group participants - these together suggest that the pattern of influence in the EC and DC groups were more effective than the implicit aspects like the number of decisions the user explored or the number of switches it made.

\section{Related Work and Conclusions}
The literature documents several experimental results on the adoption of behaviors including network structure –  such as the study conducted in \cite{centola2010spread,bao2013cumulative} and decision making \cite{grujic2012three, schobel2016social}. Under these study, it was observed that individual adoption is much more likely when participants received social reinforcement from multiple neighbors in the social network as opposed to a single exposure. However as a major contribution, they study the effect of network structure on the dynamics of behavioral diffusion. Contrary to this, we quantify influence using only the number of signals temporally sent to a user irrespective of how the signals diffused to its neighbors prior to its own adoption. Our focus here is on using social influence as a strategic tool for exploitation. \\
\textbf{Acknowledgments}. Some of the authors are supported through the ARO grant W911NF-15-1-0282.




{\footnotesize
\begin{thebibliography}{1}

\bibitem{sec_sauvik}
Das, Sauvik, et al. "The role of social influence in security feature adoption." Proceedings of the ACM conference on computer supported cooperative work \& social computing. ACM, 2015.

\bibitem{soc_inf_sec}
Hu, Qing, Paul Hart, and Donna Cooke. "The role of external and internal influences on information systems security–a neo-institutional perspective." The Journal of Strategic Information Systems 16.2 (2007): 153-172.

\bibitem{sauvik_2}
Das, Sauvik, et al. "The effect of social influence on security sensitivity." 10th Symposium On Usable Privacy and Security ({SOUPS} 2014). 2014.

\bibitem{shakarian}
Shakarian, Paulo, Jana Shakarian, and Andrew Ruef. Introduction to cyber-warfare: A multidisciplinary approach. Newnes, 2013.



\bibitem{centola2010spread}
Centola D.
\newblock The spread of behavior in an online social network experiment.
\newblock science. 2010;329(5996):1194--1197.


\bibitem{schobel2016social}
Sch{\"o}bel M, Rieskamp J, Huber R.
\newblock Social influences in sequential decision making.
\newblock PloS one. 2016;11(1):e0146536. 


\bibitem{bao2013cumulative}
Bao P, Shen HW, Chen W, Cheng XQ.
\newblock Cumulative effect in information diffusion: empirical study on a
  microblogging network.
\newblock PloS one. 2013;8(10):e76027.

\bibitem{grujic2012three}
Gruji{\'c} J, Eke B, Cabrales A, Cuesta JA, S{\'a}nchez A.
\newblock Three is a crowd in iterated prisoner's dilemmas: experimental
  evidence on reciprocal behavior.
\newblock Scientific reports. 2012;2:638.

\bibitem{survival}
Aral, Sinan, and Dylan Walker. "Creating social contagion through viral product design: A randomized trial of peer influence in networks." Management science 57.9 (2011): 1623-1639.


\bibitem{lakkaraju2015controlled}
Lakkaraju K, Medina B, Rogers AN, Trumbo DM, Speed A, McClain JT.
\newblock The Controlled, Large Online Social Experimentation Platform (CLOSE).
\newblock In: International Conference on Social Computing, Behavioral-Cultural
  Modeling, and Prediction. Springer; 2015. p. 339--344..
\end{thebibliography}
}
%

\end{document}